\documentclass[prl,twocolumn,a4paper,showpacs,superscriptaddress]{revtex4}
\usepackage{amsmath}
\usepackage{amsfonts}
\usepackage{graphicx}
\usepackage{longtable}
\newcommand{\be}{\begin{equation}}
\newcommand{\ee}{\end{equation}}

\newcommand{\ba}[1]{\left(\begin{array}{#1}}
\newcommand{\ea}{\end{array}\right)}
\begin{document}

\title{Quantum dynamical maps and Markovianity} 
\author{A. R. Usha Devi}
\email{arutth@rediffmail.com}
\affiliation{Department of Physics, Bangalore University, 
Bangalore-560 056, India}
\affiliation{Inspire Institute Inc., Alexandria, Virginia, 22303, USA.}
\author{A. K. Rajagopal} 
\affiliation{Inspire Institute Inc., Alexandria, Virginia, 22303, USA.}
\author{Sudha} 
\affiliation{Inspire Institute Inc., Alexandria, Virginia, 22303, USA.}
\affiliation{Department of Physics, Kuvempu University, Shankaraghatta, Shimoga-577 451, India.}
\date{\today}

\begin{abstract} 
It is known that the time evolution of a subsystem from an initial state to two later times, $t_1,t_2\ (t_2>t_1)$, are both completely positive (CP) but it is shown here that in the intermediate times between $t_1$ and $t_2$, in general, it need not be CP. This reveals the key to the Markov (if CP) and nonMarkov (if NCP) avataras of the intermediate dynamics. This is brought out based on $A$ and $B$ dynamical maps -- without resorting to Master equation approach. The choice of tensor product form for the global initial state points towards the system-environment interaction dynamics as the sole cause for Markovianity/non-Markovianity. A succinct summary of the results is given in the form of a table.  
\end{abstract}
\maketitle

There are three basic issues in open system quantum dynamics: (1) Role of initial correlations in system-environment state (2) Nature of the interaction between the system and its environment, and (3) Nature of the dynamics at intermediate times. The first two issues were addressed in Ref.~\cite{Jordan,Rosario,Anil, Kavan}. The purpose of this paper is to show that the third issue is central to open system evolution. This is  addressed here by making use of Choi-Jamiolkowski isomorphism~\cite{Choi, ji}. It should be noted that there is no use of master equation in this development.

Understanding the basic nature of  dynamical evolution of a quantum system, which interacts with an inaccessible environment, attracts growing importance in recent years~\cite{Breuer, Alicki}. This offers the key to achieve control over quantum systems  -- towards their applications in the emerging field of quantum computation and communication~\cite{Niel}. While the overall system-environment state evolves unitarily, the dynamics governing the system is described by a completely positive (CP), trace preserving  map~\cite{ECGS1, Choi,ji}. The concept of $A$ and $B$ dynamical maps  was first introduced --  as a quantum extension  of classical stochastic dynamics --  by Sudarshan, Mathews, Rau and Jordan (SMRJ)~\cite{ECGS1} nearly five decades ago. Its fundamental implications in open-system evolution  are being put to experimental tests nowadays. We explore here the explicit structure of dynamical maps at intermediate times to bring out the Markov or non-Markov avataras of open system evolution.

Markov approximation holds when the future dynamics  depends only on the present state    -- and not on the history of the system i.e.,  memory effects are negligible. The corresponding Markov  dynamical map constitutes a trace preserving, CP, continuous one-parameter quantum  semi-group~\cite{Lindblad,GKS}.  Markov dynamics governing the evolution of the system density matrix is conventionally described by Lindblad-Gorini-Kossakowski-Sudarshan (LGKS)  master equation~\cite{Lindblad,GKS} $\frac{d\rho}{dt}={\cal L}\rho$, where ${\cal L}$ is the time-independent Lindbladian operator generating the underlying  quantum  Markov semi-group. Generalized  Markov processes are formulated in terms of time-dependent Lindblad generators and the associated trace preserving CP dynamical map is a two-parameter divisible map~\cite{div,RHP} and this too corresponds to memory-less Markovian evolution. 
%However,  memory effects can not  be ignored in many physical processes of significance. Deviation from Markovianity could manifest itself in a 
%temporary regain of energy/information during the course of evolution of the system  --  as a result of the back-reaction of the ennvironment on the %system. 
Last few years have witnessed  intense efforts towards formulating, characterizing and quantifying non-Markovian  dynamics~\cite{div, RHP, NM,AKRUR,AKRUS, BLP,  haikka, kos, Hou}.  

NCP dynamical maps do make their presence felt in the open-system dynamics obtained from the
joint unitary  evolution --  if the system and environment are in an
initially quantum correlated state~\cite{Jordan,Rosario, Anil, Kavan}. In such cases, the open-system evolution turns out to be non-Markovian~\cite{AKRUS}. However, the source of such non-Markovianity could not be attributed entirely to either initial system-environment correlations or their dynamical interaction. This issue gets refined if initial global state is in the tensor product form, in which case the sole cause of Markovianity/non-Markovianity {\em is} dynamics. It is this aspect that is addressed here. It may be noted that these issues could not be addressed in the generalized master equation approaches~\cite{RHP, NM, BLP,  haikka, kos, Hou}.

We now begin with a brief review of the $A$ and $B$ dynamical maps~\cite{ECGS1}.
These dynamical maps  transform the initial system density matrix $\rho_S(t_0)$ to final density matrix $\rho_S(t)$ via,
\begin{eqnarray}
\label{A}
\left[\rho_S(t)\right]_{a_1a_2}&=&\sum_{a_1',a_2'} \left[A(t,t_0)\right]_{a_1a_2;a_1'a_2'}\, \left[\rho_S(t_0)\right]_{a_1'a_2'},  \\ 
\label{B}
\left[\rho_S(t)\right]_{a_1a_2}&=&\sum_{a_1',a_2'} \left[B(t,t_0)\right]_{a_1a_1';a_2a_2'}\, \left[\rho_S(t_0)\right]_{a_1'a_2'}, \\  && \ \ \ \  \ a_1,a_2,a_1',a_2'=1,2,\ldots, d  \nonumber
\end{eqnarray} 
where the realligned matrix $B$ is defined by, 
\begin{equation}
\label{bdef}
B_{a_1a_1';a_2a_2'}= A_{a_1a_2;a_1'a_2'}. 
\end{equation} 

The identity $A$-map has the following structure:
\begin{eqnarray}
\label{idmap}
\left[\rho_S(t)\right]_{a_1a_2}&=&\sum_{a_1',a_2'}\, A^{\rm Id}_{a_1a_2;a_1'a_2'}\, \left[\rho_S(t_0)\right]_{a_1'a_2'} = \left[\rho_S(t_0)\right]_{a_1a_2} \nonumber \\ & &\Rightarrow A^{\rm Id}_{a_1a_2;a_1'a_2'}=\delta_{a_1,a_1'}\, \delta_{a_2,a_2'}  
\end{eqnarray} 

The requirement that the density matrix $\rho_S(t)$  has unit trace, Hermitian and is positive semi-definite leads to the conditions~\cite{ECGS1} 
\begin{widetext} 
\begin{eqnarray}
\label{prop}
{\rm Trace\ preservation}:&&\ \ \sum_{a_1}A_{a_1a_1;a_1'a_2'}=\delta_{a_1'a_2'},\ \ \ \sum_{a_1}B_{a_1a_1';a_1a_2'}=\delta_{a_1'a_2'} ,\nonumber \\
%&& \hskip 1in \bullet\  {\rm Note\ that}\ \ {\rm Tr}[B]= \sum_{a_1,a_1'}B_{a_1a_1';a_1a_1'}=\sum_{a_1'}\delta_{a_1'a_1'}=d\nonumber \\ 
{\rm Hermiticity}:&&\ \ A_{a_1a_2;a_1'a_2'}=A^{*}_{a_2a_1;a_2'a_1'},\ \ \ B_{a_1a_1';a_2a_2'}=B^{*}_{a_2a_2';a_1a_1'} \\
%&& \hskip 1in \bullet\  {\rm We\ obtain\ }\ \ B^\dag= B\nonumber \\ 
{\rm Positivity}: && \ \ \sum_{a_1,a_2,a_1',a_2'}\,x^*_{a_1}\,x_{a_2}\, A_{a_1a_2;a_1'a_2'}\, y_{a_1'}\,y_{a_2'}^*\geq 0,\ \ \ 
\sum_{a_1,a_2,a_1',a_2'}\, x^*_{a_1}\, y_{a_1'}\, B_{a_1a'_1;a_2a_2'}\,x_{a_2}\, y_{a_2'}^*\geq 0 \nonumber 
% && \hskip 1in \bullet\  {\rm  We\ thus\ obtain\ }\ \ B\geq 0  
\end{eqnarray}
\end{widetext}
Clearly, the $B$ map is positive, Hermitian $d^2\times d^2$ matrix with trace $d$. We would also like to point out here that the composition of two 
$A$-maps, $A_1\ast A_2$ is merely a matrix multiplication, whereas it is not so in its $B$-form. 

Let us consider unitary evolution of global system-environment state $\rho_S(t_0)\otimes \rho_E(t_0)$ from an initial time $t_0$ to a final time $t_2$ -- passing through an intermediate instant $t_1$ (i.e., $t_0<t_1<t_2$). The $A$-map associated with $t_0$ to $t_1$ and that between $t_0$ to $t_2$ are readily identified as follows:
\begin{eqnarray}
\label{a1}
&&{\rm Tr}_E\left[U(t_j,t_0)\rho_S(t_0)\otimes \rho_E(t_0)U^\dag(t_j,t_0)\right]=A(t_j,t_0)\, \rho_S(t_0)\nonumber \\ 
&&\ \ \ \ \ \ \ \ \ \ = \rho_S(t_j),\ \ j=1,2.  
\end{eqnarray} 
The dynamical map $A(t_j,t_0)$ is CP. 

From the composition law $U(t_2,t_0)=U(t_2,t_1)U(t_1,t_0)$ of unitary evolution, we obtain, 
\begin{widetext}
\begin{equation}
 \label{a2}
{\rm Tr}_E\left[U(t_2,t_1)\left\{U(t_1,t_0)\rho_S(t_0)\otimes \rho_E(t_0)U^\dag(t_1,t_0)\right\}U^\dag(t_2,t_1)\right]=A(t_2,t_0)\, \rho_S(t_0).  
\end{equation} 
\end{widetext} 
If $\left\{U(t_1,t_0)\rho_S(t_0)\otimes \rho_E(t_0)U^\dag(t_1,t_0)\right\}=\rho_S(t_1)\otimes\rho_E(t_1)$ (memory-less reservoir condition) then 
the LHS of (\ref{a2}) is also given by, 
\begin{equation}
 \label{a3}
{\rm Tr}_E\left[U(t_2,t_1)\rho_S(t_1)\otimes\rho_E(t_1)U^\dag(t_2,t_1)\right]=A(t_2,t_1)\, \rho_S(t_1).  
\end{equation} 
With $j=1$ in (\ref{a1}), we substitute $\rho_S(t_0)=A^{-1}(t_1,t_0)\rho_S(t_1)$ in (\ref{a2}) to identify the intermediate $A$ map as,
\begin{equation}
\label{a4}
A(t_2,t_1)=A(t_2,t_0)A^{-1}(t_1,t_0). 
\end{equation}  
We would like to emphasize that when the environment is passive ( Markovian dynamics), the intermediate $A$-map has the composition as in (\ref{a4}). In such cases $A(t_1,t_2)$ is ensured to be CP -- otherwise it is NCP, and hence non-Markovian. When $\rho_{SE}(t_1)\neq \rho_S(t_1)\otimes \rho_E(t_1)$, implying system-environment correlations at any intermediate time $t_1$, the composition (\ref{a4}) {\em does not} hold. This construction provides us with a test of the CP  or NCP  nature of the process by explicit evaluation of the RHS of (\ref{a4}). 
(The corresponding intermediate $B$-map $B(t_2,t_1)$ is positive if the process is CP  -- otherwise it is NCP). 

We now observe that Jamiolkowski isomorphism~\cite{ji} provides an insight  that the $B$-map is directly related to a $d^2\times d^2$ system-ancilla bipartite density matrix. More specifically, 
the action of the map $A^{\rm Id}\otimes A$ on the maximally entangled system-ancilla state $\vert\psi_{\rm ME}\rangle=\frac{1}{\sqrt{d}}\sum_{i=0}^{d-1}\vert i,i\rangle$ results in $\frac{1}{d}\, B$ i.e.,
\begin{equation}
\label{th}
\left[A^{\rm Id}\otimes A \right]\vert\psi_{\rm ME}\rangle\langle\psi_{\rm ME}\vert=\frac{1}{d}\, B 
\end{equation}
gives an explicit matrix representation for the $B$-map
(Here $A^{\rm Id}$ is the identity A-map (see (\ref{idmap}), which leaves the ancilla undisturbed).

%Consider the action of the dynamical map $A^{\rm Id}\otimes A$ (where $A$ is the dynamical map defined in Eq.(1) and $A^{\rm Id}$ is the identity map %(\ref{idmap}) on the  maximally entangled bipartite state $\vert\psi_{\rm ME}\rangle.$  
In detail, we have,
 \begin{widetext}
 \begin{eqnarray}
\label{Jam}
\sum_{a'_1,a'_2,b'_1,b'_2}\,\left[A^{\rm Id}\otimes A \right]_{a_1b_1a_2b_2;a'_1b'_1a'_2b'_2}\, 
\left[\vert\psi_{\rm ME}\rangle\langle\psi_{\rm ME}\vert\right]_{a_1'b_1';a_2'b_2'}
&=& \frac{1}{d}\, \sum_{a'_1,a'_2,b'_1,b'_2}\, \delta_{a_1,a'_1}\delta_{a_2,a'_2}\, A_{b_1b_2;b_1'b_2'}\, \delta_{a_1',b_1'}\delta_{a_2',b_2'}\nonumber \\ 
&=& \frac{1}{d}\,  A_{b_1b_2;a_1a_2}=\frac{1}{d}\, B_{b_1a_1;b_2a_2}. 
\end{eqnarray}
\end{widetext}
% where  we have employed the expression for the the identity map on the first system (see Eq.(\ref{idmap})) and used
% $\left[\vert\psi_{\rm ME}\rangle\langle\psi_{\rm ME}\vert\right]_{a_1'b_1';a_2'b_2'}=\frac{1}{d}\sum_{i,j=1}^{d}\,\langle a_1'b_1' \vert %i,i\rangle\langle j,j\vert a_2'b_2'\rangle = \frac{1}{d}\, \delta_{a_1',b_1'}\delta_{a_2',b_2'}.$

In other words, Jamiolkowski isomorphism maps {\em every completely positive dynamical map} acting on $d$ dimensional space to 
a positive definite $d^2\times d^2$ bipartite density matrix $\rho_{ab}$ (which is just $\frac{1}{d}\, B$)   -- whose partial trace (over the first subsystem index) is a maximally disordered state. This result is powerful -- as one may now identify several toy models of dynamical $B$ maps  to investigate the nature of dynamics, which will be presented next.  

Consider the two-qubit (system-ancilla) density matrix  $\rho_{ab}(t)=\frac{[1-p(t)]}{4}\, I_2\otimes I_2 +p(t)\, \vert\psi_{\rm ME}\rangle\langle\psi_{\rm ME}\vert$  -- with $0\leq p(t) \leq 1$ being a suitable function of time, and $\vert\psi_{\rm ME}\rangle=\frac{1}{\sqrt{2}}\left(\vert 0_a, 0_b\rangle+\vert 1_a,1_b\rangle\right)$  -- as a prototype of our dynamical $B$ map. For a dynamical map, time dependence in $p(t)$ occurs due to the  underlying Hamiltonian evolution.   This state  
 is especially important in that it exhibits both separable and entangled states, as its characteristic parameter $p(t)$ is varied. Its use here as a valid $B$-map is novel in identifying Markovianity of dynamics.
%Since, $\vert\psi_{\rm ME}\rangle=\frac{1}{\sqrt{2}}\left(\vert 0_a, 0_b\rangle+\vert 1_a,1_b\vert\rangle\right)$ and as $B=2\rho_{ab}$,  we have 

On evaluating the corresponding $A$ map $A(t,0)$, one can obtain the intermediate dynamical map 
$A(t_2,\,t_1)=A(t_2,\,0)A^{-1}(t_1,\,0)$. The intermediate time $B$-map $B(t_2,t_1)$ is then given by 
\be
B(t_2,\,t_1)=\frac{[p(t_1)-p(t_2)]}{2p(t_1)}\, I_2\otimes I_2+\frac{2p(t_2)}{p(t_1)}\, \vert\psi_{\rm ME}\rangle\langle\psi_{\rm ME}\vert.
\ee
Its eigenvalues are $\lambda_1=\lambda_2=\lambda_3=\frac{p(t_1)-p(t_2)}{2p(t_1)}$ and $\lambda_4=\frac{p(t_1)+3p(t_2)}{2p(t_1)}$.  
A choice $p(t)=\cos^{2N}(a t)$ for any $N\geq 1$ leads to NCPness of the intermdiate map  -- as  the eigenvalues   $\lambda_{1,\,2,\,3}$ of $B(t_2,t_1)$  turn out to be negative -- and hence  non-Markovian  dynamics ensues. Another choice  $p(t)=e^{-\alpha t}$ corresponds to a CP intermediate map -- resulting in a Markovian process. In this case, we also find that $A(t_2,t_1)=A(t_2-t_1)$ and this forms a Markov semigroup. 
However, if $p(t)=e^{-\alpha\, t^\beta},\ \ (\beta\neq 1)$, the intermediate map is still CP (and hence Markovian),  though  $A(t_2,t_1)\neq A(t_2-t_1)$ -- and hence it does not constitute a one-parameter semigroup.   

Another important feature that we wish to illustrate through this example is the following: Concurrence  of $\rho_{ab}(t)=\frac{1}{d} \, B(t,0)$ 
(given by $C=\frac{3p(t)-1}{2}$) can never increase as a result of  Markovian evolution. This is because ensuing dynamics is a  local  CP map on the system. Any temporary regain of system-ancilla entanglement during the course of evolution is clearly attributed to the back-flow from environment to the system -- which is a signature of non-Markovian process. This feature is displayed in Fig.~1 by plotting the concurrence of $\rho_{ab}(t)$ for different choices of $p(t)$.  

\begin{figure}[h]
\includegraphics*[width=2.5in,keepaspectratio]{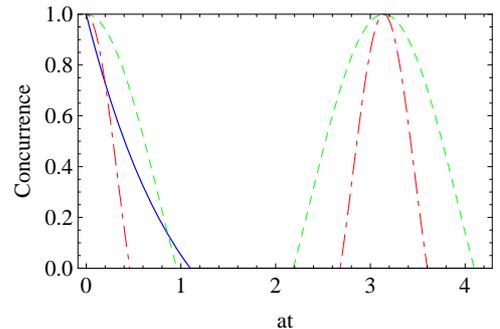}
\caption{Concurrence $C=\frac{3p(t)-1}{2}$ of the system-ancilla state $\rho_{ab}(t)=\frac{[1-p(t)]}{4}\, I_2\otimes I_2 +p(t)\, \vert\psi_{\rm ME}\rangle\langle\psi_{\rm ME}\vert$, vs scaled time $at$, for the following choices (i) Markov process: $p(t)=e^{-at}$ (solid line) and (ii) non-Markov process:  $p(t)=\cos^{2N}(a\, t),\ N=1$ (dashed line) and $N=5$ (dot-dashed line). Note that there is a death and re-birth of entanglement (dash, dot-dashed lines) due of  back-flow from environment.}   
\end{figure}

We now present two  Hamiltonian models, which give rise to explicit structure of time dependence in the open system evolution.  Interaction Hamiltonian considered here is~\cite{Hou} 
\be
H=\frac{A}{\sqrt{N}}\sigma_z \sum_{k=1}^N \, \sigma_{k\,z}.
\ee                             
This is a simplified model of a hyperfine interaction of a spin-1/2 system with $N$ spin-1/2 nuclear environment in a quantum dot.  
Taking the initial system-environment state to be $\rho_S(0)\otimes \frac{I_{2^N}}{2^N}$, the dynamical $A$-map is obtained by evaluating 
${\rm Tr}_E\left[U(t,0)\,\rho_S(0)\otimes \frac{I_{2^N}}{2^N}\, U^\dag(t,0)\right]$ (where $U(t,0)={\rm Exp}[-i\,H\, t]$): 
\begin{eqnarray}
A(t,0)&=&\frac{1}{2}\left(1-x(t)\right)\sigma_z\otimes\sigma_z+\frac{1}{2}\left(1+x(t)\right)I_2\otimes I_2 \nonumber \\
& & x(t)=\cos^N \left (\frac{2At}{\sqrt N}) \right). 
\end{eqnarray}
From this, the intermediate map $A(t_2,t_1)$ (see (\ref{a4})) and in turn the corresponding $B(t_2,t_1)$ may be readily evaluated. We obtain, 
\begin{eqnarray} 
B(t_2,\,t_1)&=&\frac{1}{2}\, \left(I_2\otimes I_2+\sigma_z\otimes\sigma_z\right)\nonumber \\ && +\frac{x(t_2)}{2x(t_1)}\left(\sigma_x\otimes\sigma_x-\sigma_y\otimes\sigma_y\right).
\end{eqnarray}  
The eigenvalues of $B(t_2,t_1)$ are $0,0,1\pm\frac{x(t_2)}{x(t_1)}.$ Clearly, the intermediate time dynamics is NCP -- and hence the process is non-Markovian.  

We now consider the open system dynamics arising from the  unitary evolution~\cite{Jordan} 
\begin{eqnarray}
U(t,0)&=&e^{-i\,t\, [\omega\, \sigma_{z}\otimes \sigma_{x}]} \\
&=&\cos(\omega\, t/2)\, I_2\otimes I_2-i\sin(\omega\ t/2)\, \sigma_z\otimes \sigma_x \nonumber
\end{eqnarray}
on the system-environment initial state $\rho_{SE}(0)=\rho_S(0)\otimes \rho_E(0)=\frac{1}{2}\left(I_2+\sigma_x\right)\otimes\frac{1}{2}\left(I_2+\sigma_z\right).$ The $A(t,0)$ map is given by, 

\be
A(t,0)=\frac{1}{2}\left(1+\cos(\omega\, t)\right)\, I_2\otimes I_2+\frac{1}{2}\left(1-\cos(\omega\, t)\right)\, \sigma_z\otimes\sigma_z. 
\ee
Following (\ref{a4}), we obtain
\begin{eqnarray}
B(t_2,t_1)&=&\frac{1}{2}\, \left(I_2\otimes I_2+\sigma_z\otimes\sigma_z\right)\nonumber \\ && +\frac{\cos(\omega\,t_2)}{2\cos(\omega t_1)}\left(\sigma_x\otimes\sigma_x-\sigma_y\otimes\sigma_y\right).
\end{eqnarray}  
The eigenvalues of the $B$-map being $0,\ 0, 1\pm \frac{\cos \omega t_2}{\cos \omega t_1}$ bringing out the NCP nature of the intermediate map  explicitly -- implying non-Markov nature of the dynamical process. This model, with initially correlated states,  has been explored in Ref.~\cite{Jordan, AKRUS} and the dynamical map turned out be NCP throughout (not merely in the intermediate time interval).

\begin{widetext}
\begin{table}
\begin{tabular}{|c|c|c|c|c|}
\hline
Initial state & $B(0,t_1)$ & $B(0,t_2)$ & $B(t_2,t_1)$ & Nature of dynamics \\
\hline 
$\rho_{SE}=\rho_S\otimes \rho_E$ & CP & CP & CP & Markov \\ 
\hline 
$\rho_{SE}=\rho_S\otimes \rho_E$ & CP & CP & NCP & non-Markov \\
\hline 
Quantum correlated $\rho_{SE}$ & NCP & NCP & -- & non-Markov \\  
\hline
\end{tabular}
\caption{CP/NCP nature of the maps  and the associated Markov/non-Markov features}
\end{table}
\end{widetext}

In conclusion, a few remarks on a 
variety of definitions of non-Markovianity in the recent literature may be recalled here.  Mainly the focus has been towards capturing the violation of semi-group property~\cite{AKRUR,AKRUS} or  more recently -- its two-parameter generalization viz the divisibility of the dynamical map~\cite{div,RHP}. Yet another measure, where non-Markovianity~\cite{BLP} is attributed to  increase of distinguishability of any pairs of states (as a result of the partial back-flow of information from the environment  into the system) and is quantified in terms of trace distance of the states. It has been shown that the two different measures of non-Markovianity  -- one based on the divisibility of the dynamical map~\cite{RHP}  and the other based upon the distinguishability of quantum states~\cite{BLP} -- need not agree with each other~\cite{haikka}.    A modified version of the criterion of Ref.~\cite{RHP} was proposed recently~\cite{Hou}. In this paper it is shown that the intermediate time evolution of the system is not necessarily CP and holds the key to the memory of the initial state in a subtle way. The results are summarized in Table~1.

%\noindent{\bf Conclusions} In this paper it is shown the intermediate time evolution of the subsystem is not necessarily CP and holds the key to the %memory of the initial state in a subtle way.


\begin{thebibliography}{0} 
\bibitem{Jordan} T. F. Jordan, A. Shaji, and E. C. G. Sudarshan, \pra {\bf 70}, 052110 (2004).
\bibitem{Rosario} C. A. Rod{\' r}guez-Rosario and E. C. G. Sudarshan, e-print arXiv:0803.1183 [quant-ph].
\bibitem{Anil} C. A. Rodr{\' i}guez-Rosario, K. Modi, A. Kuah, A. Shaji, and
E. C. G. Sudarshan, J. Phys. A {\bf 41}, 205301 (2008).
\bibitem{Kavan} K. Modi and E. C. G. Sudarshan, \pra {\bf 81}, 052119 (2010).
\bibitem{Choi} M. D. Choi, Can. J. Math. {\bf 24}, 520 (1972); Linear Algebra and Appl. {\bf 10}, 285 (1975).
\bibitem{ji} A. Jamiolkowski, Reports on Mathematical Physics, {\bf 3}, (1972).  
\bibitem{Alicki} R. Alicki and K. Lendi, {\em Quantum Dynamical Semigroups
and Applications} (Springer, Berlin, 1987).
\bibitem{Breuer} H.-P. Breuer and F. Petruccione, {\em The Theory of Open
Quantum Systems} (Oxford Univ. Press, Oxford, 2007). 
\bibitem{Niel} M. A. Nielsen and I. L. Chuang, {\em Quantum Computation
and Quantum Information} (Cambridge Univ. Press, Cambridge, 2000).
\bibitem{ECGS1} E. C. G. Sudarshan, P. Mathews, and J. Rau, Phys. Rev. {\bf 121}, 920 (1961); T. F. Jordan and E. C. G. Sudarshan, J. Math. Phys. {\bf 2}, 772 (1961).
\bibitem{Lindblad} G. Lindblad, Comm. Math. Phy. {\bf 48}, 119 (1976). 
\bibitem{GKS} V. Gorini, A. Kossakowski, and E. C. G. Sudarshan, J. Math. Phys. {\bf 17}, 821 (1976).
\bibitem{div} M. M. Wolf, J. Eisert, T. S. Cubitt and J. I. Cirac, Phys. Rev. Lett. 101, 150402 (2008).
\bibitem{RHP} A. Rivas, S. F. Huelga and M. B. Plenio, \prl {\bf 105}, 050403 (2010). 
\bibitem{NM} H.-P. Breuer, \pra {\bf 69} 022115 (2004);
{\it ibid.} {\bf 70}, 012106 (2004); S. Daffer, K. W{\' o}dkiewicz, J. D. Cresser, and J. K. McIver, \pra {\bf 70}, 010304
(2004); H.-P. Breuer and B. Vacchini,\prl {\bf 101} (2008) 140402; \pre
{\bf 79}, 041147 (2009); A. Kossakowski and R. Rebolledo, Open Syst. Inf. Dyn.
{\bf 14}, 265 (2007); {\bf 15}, 135 (2008); {\bf 16}, 259 (2009); D. Chru{\' s}ci{\' n}ski and A. Kossakowski,
\prl {\bf 104}, 070406 (2010)
\bibitem{AKRUR} A. K. Rajagopal, A. R. Usha Devi and R. W. Rendell, \pra {\bf 82}, 042107 (2010).
\bibitem{AKRUS} A. R. Usha Devi, A. K. Rajagopal, Sudha, \pra {\bf 83}, 022109 (2011).  
\bibitem{BLP} H.-P. Breuer, E.-M. Laine, J. Piilo, Phys. Rev.Lett. 103, 210401 (2009); E.-M. Laine, J. Piilo, H.-P. Breuer, \pra {\bf 81}, 062115 (2010).
\bibitem{haikka} P. Haikka, J.D. Cresser and S. Maniscalco \pra {\bf 83}, 012112 (2011). 
\bibitem{kos} D. Chru{\' s}ci{\' n}ski,  A. Kossakowski and A. Rivas, eprint arXiv: 1102.4318v2
\bibitem{Hou} S. C. Hou, X. X. Yi, S. X. Yu and C. H. Oh, eprint arXiv: 1102.4659. 
\end{thebibliography}
\end{document}